\documentclass{elsart}
\usepackage{graphicx}

\begin{document}

\newcommand{\beq}{\begin{equation}}
\newcommand{\eeq}{\end{equation}}
\newcommand{\beqn}{\begin{equation}}
\newcommand{\eeqn}{\end{equation}}
\newcommand{\mrm}{\mathrm}
\newcommand{\mbf}{\mathbf}
\newcommand{\ol}{\overline}
\newcommand{\fr}{\frac}
\newcommand{\kT}{k_\mrm{B}T}
\newcommand{\kB}{k_\mrm{B}}

\begin{frontmatter}

\title{
 Reversible mesoscopic model of protein adsorption:
 From equilibrium to dynamics
}

\author[ELTE]{G. J. Sz\"oll\H{o}si\corauthref{cor}},
\corauth[cor]{ssolo@angel.elte.hu}
\author[ELTE]{I. Der{\'e}nyi} and
\author[LSST]{J. V\"or\"os}

\address[ELTE]{
 Department of Biological Physics, E\"otv\"os University,
 P\'azm\'any P.\ stny.\ 1A, H-1117 Budapest, Hungary
}
\address[LSST]{
 BioInterface Group, Laboratory for Surface Science and Technology,
 ETH Z\"urich,
 Wagistrasse 2, 8952 Schlieren, Switzerland
}

\begin{abstract}
We present a thermodynamically consistent mesoscopic model of protein
adsorption at liquid-solid interfaces. First describing the equilibrium
state under varying protein concentration of the solution and binding
conditions, we predict a non-trivial (non-monotonic) dependence of the
experimentally observable properties of the adsorbed layer (such as the
surface density and surface coverage) on these parameters. We
subsequently proceed to develop a dynamical model consistent with the
equilibrium description, which qualitatively reproduces known
experimental phenomena and offers a promising way of studying the
exchange of the adsorbed proteins by the proteins of the solution.
\end{abstract}


\end{frontmatter}


\section{Introduction}

Protein adsorption at liquid-solid interfaces is a fundamental problem
of several diverse areas of biotechnology. A short list would include,
biocompatibility of implants, blood clotting, filter fouling and
protein chip technology. It is also a problem of general theoretical
interest, especially because despite the considerable body of
experimental data and empirical phenomena -- due in part to the very
precise methods available for the measurement of the adsorbed amount --
the literature is thus far lacking a mesoscopic (particle level)
description that can account for all of the experimental observations
\cite{Calonder-VanTassel'01}.
Below we present a model that will -- we hope -- not only meet the
above criteria, but do so in a physically consistent manner.

The basic treatise behind various particle level descriptions of
protein adsorption is that adsorbed proteins can be in several
different states of different surface (or footprint) sizes. This is
supported by experimental evidence that adsorbed proteins undergo
surface-induced conformational change
\cite{Norde-rev'95,Bujis-Norde'96,Wertz-Santore'99}
characterized by a substantial growth of the surface contact area.
Combined with evidence
\cite{Norde-Favier'95}
that under conditions of high surface coverage these transitions are
sterically hindered by neighboring proteins on the surface, the usual
mesoscopic description is completed by assuming a hard-core repulsive
interaction between proteins -- both on the surface and in the
solution.

Although proteins adsorb to a two-dimensional (2D) substrate, we will
further use a one-dimensional (1D) approximation. This simplification
is justified, on one hand, by the fact that there is no thermodynamic
phase transition (which is usually sensitive to the dimensionality) in
the adsorption process, and on the other hand, because the main
features of adsorption (such as jamming, slow convergence toward
equilibrium, short-range spatial correlations) are independent of the
spacial dimensions. 1D models already have the ingredients that make
their behavior non-trivial and qualitatively similar to that observed
in higher dimensions, and they are much easier to treat analytically
\cite{RSA-rev}.

Surface-protein systems of interest typically exhibit tightly bound
protein states with binding energies upto several tens of $\kT$
\cite{Norde'92}.
This is the reason that most models to date have incorporated at least
partial irreversibility to explain the obvious deviation of protein
adsorption from simple Langmuir-like behavior
\cite{Norde'00}.
The classical irreversible model of particle adsorption, random
sequential adsorption (RSA) (for a recent review see
\cite{RSA-rev}),
in which adsorption is considered to be permanent and the adsorbed
system of particles on the surface is eventually described by a jammed
state with no more free space available for particle deposition, is
inherently incapable of describing any process involving desorption of
proteins. The second class of models
\cite{Lundstrom'85,VanTassel-Voit'96,Wertz-Santore'02n1,Wertz-Santore'02n2,VanTassel'03}
differentiate between reversibly adsorbed states of small size and
irreversibly adsorbed larger states. These models rely on the
assumption that the escape from the strongly bound states is
practically impossible under reasonable experimental time scales. This
notion is supported by the experimental evidence of long relaxation
times of the observed properties. We show that slow relaxation of
experimental observables (typically surface density) can be equally
well described by a reversible model, and that such an approach has a
non-trivial equilibrium state distribution, with several unprecedented
predictions for the concentration and binding strength dependence of
the surface density and the surface coverage, respectively. We also
argue that exchange effects, widely observed for non-biological
polymers
\cite{Fleer'93},
and to a limited extent for proteins
\cite{Ball'03}
can only be treated in the context of a reversible model such as ours.

\begin{figure}
\begin{center}
\includegraphics[scale=1]{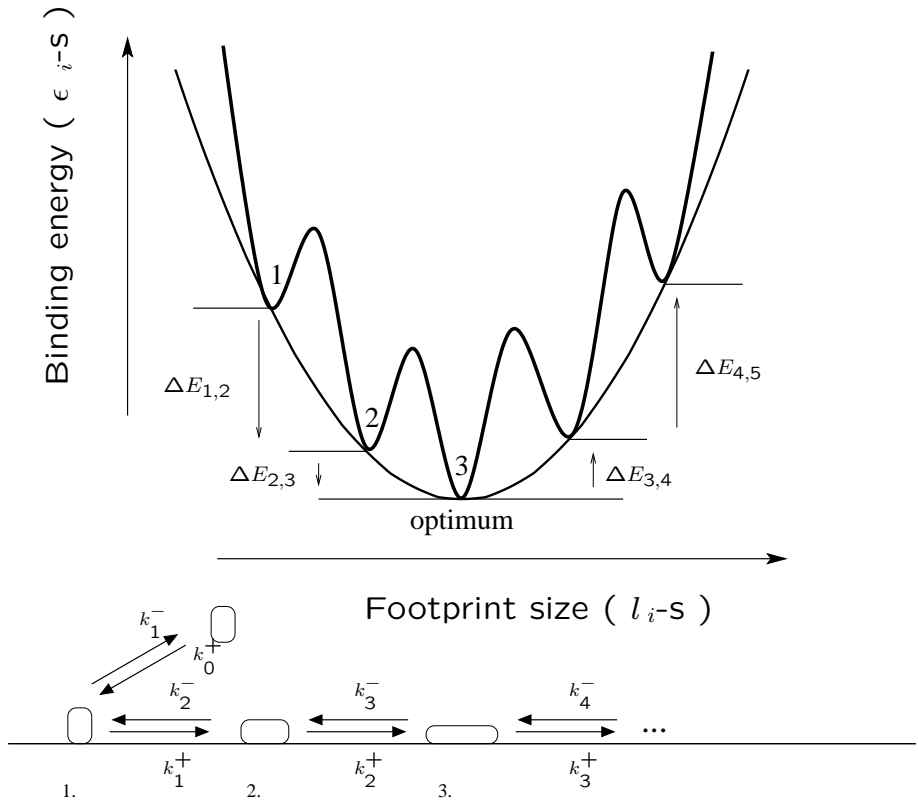}
\end{center}
\caption{
 Adhesion potential of a protein, where each of its states $i$
(characterized by a binding energy $\varepsilon_i$ and a footprint size
$l_i$) is separated from the neighboring states by potential barriers.
The corresponding transition rates are $k_i^{\pm}$ (defined in Sec.\
\ref{secdyn}).
}
\label{potfig}
\end{figure}


\section{Equilibrium model}

In our model we consider a 1D line of adsorption of size $L$ in contact
with a protein solution of constant concentration $c$. Each proteins
being adsorbed on this line is considered to be in one of $M$ different
states, each characterized by an $\varepsilon_i$ binding energy ($1\leq
i\leq M$) and a monotonically increasing\footnote{
  we can assume $\{l_i\}$ to be
  monotonically increasing as for any set of $\{l_i\}$ and
  $\{\varepsilon_i\}$ we can obtain this by appropriately rearranging
  the $\{l_i\}$-s, because there is no constraint on the
  $\{\varepsilon_i\}$-s.}
$l_i$ linear footprint size (see Fig.\ \ref{potfig}). Proteins in the
smallest footprint state ($i=1$) can undergo adsorption from and
desorption to the protein solution, characterized by a $k_\mrm{a}$
adsorption and a $k_\mrm{d}$ desorption rate. Adsorption and desorption
in other states are neglected, because these states have lower energies
on the surface and higher energies in the solution than those of the
smallest footprint state, and thus, represent a much slower
adsorption-desorption kinetics. Proteins interact via a hard-core
repulsion, i.e., they are not allowed to overlap.

Given a set of values for the parameters
($L,c,\{\varepsilon_i\},\{l_i\},k_\mrm{a},k_\mrm{d}$) the model has a
unique equilibrium distribution of states $\{n_i\}$, where $n_i$
denotes the number of adsorbed proteins in state $i$, and
\beq \sum_{i=1}^M n_i = N \label{N}\eeq
is the total number of proteins adsorbed on the surface. The
distribution must obey the
\beq \sum_{i=1}^M n_i l_i \le L \label{L_constraint}\eeq
spatial constraint. Our reversible dynamical models (defined later)
will be explicitly shown to converge to this equilibrium distribution.


\subsection{The conditional free energy}

The above equilibrium distribution can be obtained by
determining the conditional free energy
\beq F(\{n_i\}) = E(\{n_i\}) - T S(\{n_i\}) - \mu N(\{n_i\})
\label{F1}\eeq
for any given distribution $\{n_i\}$ (obeying the spatial constraint),
and finding the distribution $\{n_i^{\mrm{min}}\}$ which minimizes it.
The energy $E(\{n_i\})$ and the entropy $S(\{n_i\})$ can be expressed
as
\beqn
E(\{n_i\}) = \sum_{i=1}^M \varepsilon_i n_i
\quad \mathrm{and} \quad
S(\{n_i\}) = - \kB \ln C(\{n_i\})
\eeqn
and $N$ is given by Eq. (\ref{N}). Now we need to count the number of
possible arrangements (configurations) $C(\{n_i\})$ of the proteins on
the surface, and express the chemical potential $\mu$ of the proteins
in the solution in terms of their concentration and
adsorption/desorption rates. Restricting ourselves to a 1D model has
the advantage that the number of configurations $C(\{n_i\})$ can be
calculated in a straightforward manner.


\subsubsection{Adsorption on a lattice}

Representing the surface as a 1D lattice of period $\delta$ (with
$L/\delta$ sites), the problem of calculating $C(\{n_i\})$ reduces to a
simple combinatoric exercise. Denoting the total occupied surface as
\beqn l = \sum_{i=1}^M n_i l_i ,\eeqn
and the number of unoccupied lattice sites as
\beq n_0 = \fr{L - l}{\delta}, \label{n0}\eeq
the number of different permutations of the $N$ adsorbed proteins and
the $n_0$ empty sites is $(n_0 + N)!$. Since proteins in the same
state, as well as the empty sites are indistinguishable, this has to be
divided by $n_0 ! \prod_{i=1}^M n_i !$ to get the number of different
configurations:
\beq C(\{n_i\}) = \frac{(n_0 + N)!}{n_0 ! \prod_{i=1}^M n_i !}
.\label{C_d}\eeq

Considering the lattice constant $\delta$ small compared to the protein
footprint size implies that the number of empty sites is much larger
than $N$. Consequently, in the above formula we can make the
approximation:
\beq \fr{(n_0 + N) !}{n_0 !} \simeq n_0^N ,\label{apC}\eeq
which is shown to be exact as $\delta \to 0$, in the Appendix.

Using Stirling's formula and neglecting the sub-linear terms in $N$:
\beqn
\ln C(\{n_i\}) \simeq
 \left( N \ln \frac{L-l}{\delta} - \sum_{i=1}^M n_i \ln n_i + N
\right).\eeqn

The entropy can now be written as
\beqn S(\{n_i\}) = \kB N \ln \frac{L-l}{\delta} - \kB \sum_{i=1}^M n_i \ln
n_i + \kB N .\eeqn

Introducing the state densities
\beqn
\rho_i = \fr{n_i}{L}
\quad \mrm{and} \quad
\rho = \sum_{i=1}^{M} \rho_i = \fr{N}{L}
\eeqn
for proteins in state $i$ and for the total adsorbed proteins,
respectively, as well as the corresponding fractional surface coverages
\beqn
\lambda_i = \fr{l_i n_i}{L}
\quad \mrm{and} \quad
\lambda = \sum_{i=1}^{M} \lambda_i = \sum_{i=1}^{M} \rho_i l_i = \fr{l}{L}
,\eeqn
we can explicitly express the extensivity of the entropy or
equivalently the lack of dependence of the entropy density $s=S/L$ on
the system size $L$, by writing the entropy density as
\beq s(\{\rho_i\}) = \fr{S(\{L\rho_i\})}{L} = \kB \left( \rho \ln
 (1-\lambda) - \rho \ln \delta - \sum_{i=1}^M \rho_i \ln
 \rho_i + \rho \right)
.\label{S_d}\eeq

The chemical potential can be derived from the requirement that for a
system in equilibrium detailed balance must hold. In our case this
requirement prescribes that the rate of adsorption to a particular site
per unit time, $\delta c k_\mrm{a}$, and the rate of desorption from
the same state, $k_\mrm{d}$, satisfy
\beqn \exp({\fr{\mu_{\delta} - \varepsilon_1}{\kT}}) =
\fr{\delta c k_\mrm{a}}{k_\mrm{d}} \eeqn
yielding
\beq \mu_{\delta} = \underbrace{\underbrace{
\varepsilon_1 + \kT \ln \frac{k_\mrm{a}}{k_\mrm{d}} }_{\mu_0}
+ \kT \ln c}_{\mu} + \kT \ln \delta =
\mu + \kT \ln\delta ,\label{mu_d}\eeq
where we have separated $\mu$ into $\mu_0$ (which depends only on the
binding conditions) and $\kT \ln c$ (which depends on the
concentration). For the sake of brevity we will consider $\kT = 1$
throughout the paper. We can then write the conditional free energy
density using (\ref{F1}), (\ref{S_d}) and (\ref{mu_d}) as
\beq \phi(\{\rho_i\}) = \fr{F(\{L\rho_i\})}{L} = \sum_{i=1}^M\rho_i
\varepsilon_i - \rho (1 + \mu) - \rho
\ln (1 - \lambda) + \sum_{i=1}^M \rho_i \ln \rho_i
,\label{phi}\eeq
where the $\ln \delta$ terms have canceled, and hence, the free energy
density is independent of the spatial resolution $\delta$.


\subsection{Equilibrium state distribution}

The free energy density (\ref{phi}) can be formally extremized by
solving
\beq
\frac{\partial \phi(\{\rho_i\})}{\partial \rho_i} =
\varepsilon_i - \mu + \ln \rho_i - \ln (1-\lambda) +
\frac{\rho}{1-\lambda} l_i= 0.
\label{min}\eeq
This yields for the equilibrium state density distribution
\beq \rho_j = (1- \lambda)
\e^{\mu} \e^{-(\varepsilon_j + \frac{\rho}{1-\lambda} l_j) },
\label{rhoj}\eeq
which must satisfy the self-consistency criteria:
\beq
\rho = \sum_{i=1}^{M} \rho_i
\quad \mrm{and} \quad
\lambda = \sum_{i=1}^M l_i \rho_i
.\label{scc}\eeq

To numerically solve the above we can proceed by defining the functions
\beqn
\rho^*(\rho,\lambda) = \sum^{M}_{i=1} \rho_i^*(\rho,\lambda)
\eeqn
and
\beqn
\lambda^*(\rho,\lambda) = \sum^{M}_{i=1} l_i \rho_i^*(\rho,\lambda)
,\eeqn
where
\beqn \rho_i^*(\rho,\lambda) =
(1- \lambda) \e^{\mu} \e^{-(\varepsilon_i + \frac{\rho}{1-\lambda} l_i) }
.\eeqn
For a given set of values of the parameters
($c,\{\varepsilon_i\},\{l_i\},k_\mrm{a},k_\mrm{d}$),
\beqn
\rho^*(\rho,\lambda) - \rho =0
\quad \mrm{and} \quad
\lambda^*(\rho,\lambda) - \lambda =0
\eeqn
define two curves in the $(\rho,\lambda)$-space, an intersection of
which (at the point $\rho = \rho^{\mrm{ext}}$ and $\lambda =
\lambda^{\mrm{ext}}$) gives a solution of Eq. (\ref{min}). Provided
that $\rho^{\mrm{ext}} \in [0,\fr{1}{\min\{l_i\}}]$ and
$\lambda^{\mrm{ext}} \in [0,1]$, this corresponds to an extremum of the
conditional free energy (\ref{phi}),
\beq
\rho_i^{\mrm{ext}} = (1- \lambda^{\mrm{ext}}) \e^{\mu} \e^{-(\varepsilon_i +
\frac{\rho^{\mrm{ext}}}{1-\lambda^{\mrm{ext}}} l_i) }
.\label{sol}\eeq
From physical considerations it is obvious that there must be at least
one such solution, which is furthermore a minimum of (\ref{phi})
(otherwise we would be left without a stable equilibrium state).
Although a formal proof is so far lacking, numerical results indicate
that there are no values of the parameters for which there would be
more than one solution.


\subsection{Linear approximation}

\begin{figure}
\begin{center}
\includegraphics[scale=0.4]{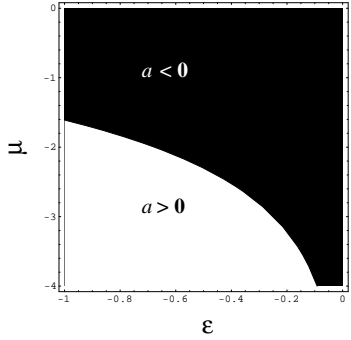}
\end{center}
\caption{
 Plot in $(\varepsilon,\mu)$-space of the $a=0$ contour for $M=5$
and $l_i=i$.
}
\label{afig}
\end{figure}

If we only wish to gain information on whether, for a particular
system, the proteins adsorbed on the surface spread out completely
(energy dominated regime), or whether they remain compact despite the
availability of energetically favorable but larger states (entropy
dominated regime), we can proceed by considering a simple linear
approximation for the binding energies as a function of the footprint
size, and neglect all higher order terms:
\beq\varepsilon_i = \varepsilon l_i
.\label{linapo}\eeq
It is immediately apparent that Eq. (\ref{sol}) specifies an
exponential form for the equilibrium state density distribution:
\beqn
\rho_j =
\underbrace{(1- \lambda) \e^{\mu}}_{b} \e^{\overbrace{-(\varepsilon +
\frac{\rho}{1-\lambda})}^{a} l_j} = b \e^{a l_j}
.\eeqn
This result of an exponential state density distribution then,
depending on the exponent $a$, corresponds either to a system -- for
negative $a$ -- where proteins remain compact compared to their
energetically more favorable (assuming $\varepsilon <0$) spread out
state, or -- for positive $a$ -- to a system where they spread out and
predominantly occupy their lowest energy state. Within the context of
our approximation the crossover between these two modes of behavior can
be analytically derived. Since for $a=0$,
$\rho_i = b$ the self-consistency criteria (\ref{scc}) become
independent of the $\rho_i$-s:
\beqn
\rho = \sum_{i=1}^{M} \rho_i = b M
\quad \mrm{and} \quad
\lambda = \sum_{i=1}^M \rho_i l_i = b \sum_{i=1}^M l_i \; .
\eeqn
Substituting for $\rho$ in the definition
$a=-(\varepsilon+\fr{\rho}{1-\lambda})$ and setting it equal to zero
we obtain
\beqn(1-\lambda) \varepsilon = - b M .\eeqn
Using the definition $b=\e^{\mu}(1-\lambda)$ we arrive at
\beq \e^{\mu} = - \fr{\varepsilon}{M}
.\label{a0}\eeq
We can see that for a given value of $\varepsilon<0$ there is always a
value of $\mu$ and consequently $c$ below which the proteins spread out
and the system is characterized by a low surface density but high
surface coverage ($a>0$), and above which the surface density is high
but the proteins are relatively loosely adhered ($a<0$) (see Fig.\
\ref{afig} as an example).

\begin{figure}
\begin{center}
\includegraphics[scale=0.4]{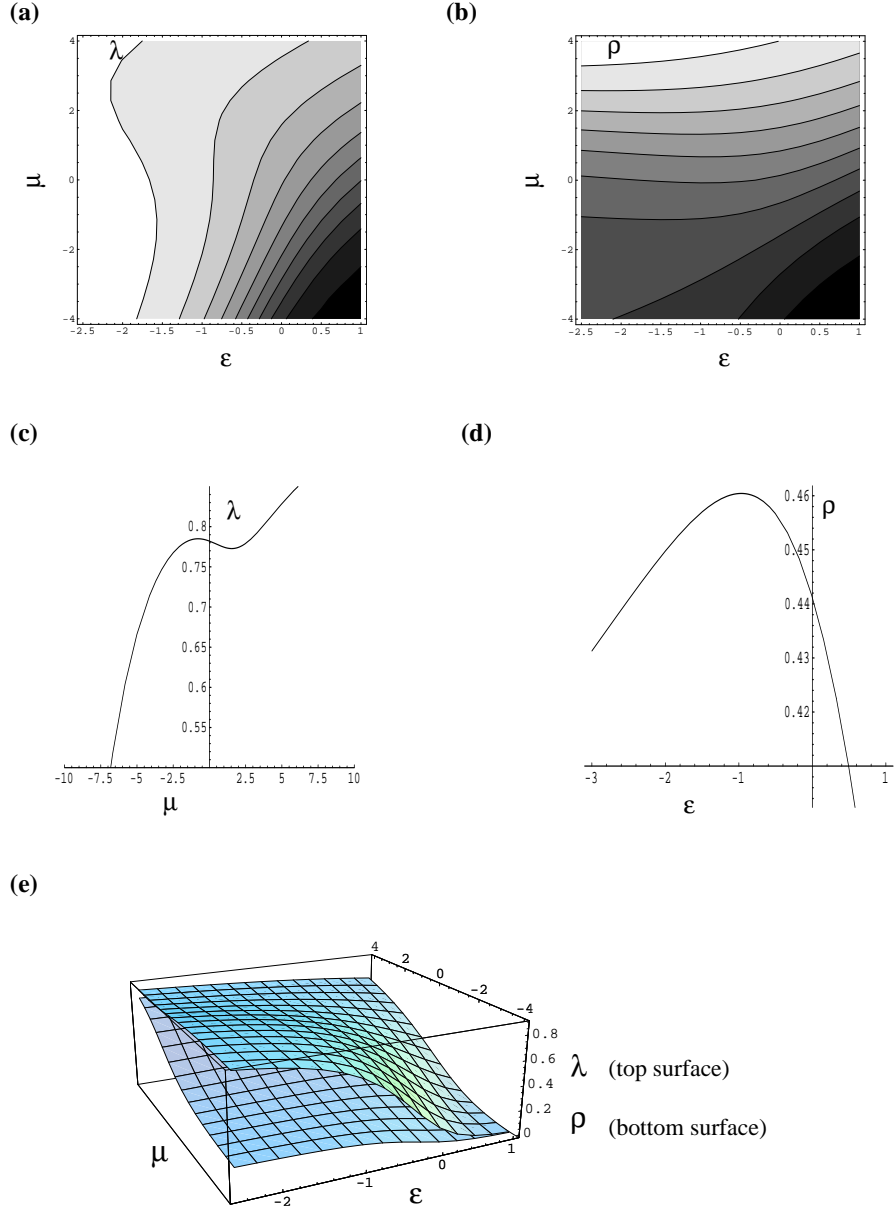}
\end{center}
\caption{
 (a) and (b) contours of $\lambda$ and $\rho$ in the
$(\varepsilon,\mu)$-space,
 (c) and (d) slices of $\lambda$ along $\mu$ at $\varepsilon=-1.2$ and
$\rho$ along $\varepsilon$ at $\mu=1$,
 (e) surface plot of $\lambda$ and $\rho$ together,
for $M = 5$ and $l_i=i$ throughout.
}
\label{RLfig}
\end{figure}

Considering the number and size of the subsequent footprints $\{l_i\}$
immutable, the equilibrium state of the adsorbed proteins can be
completely determined for any given value of $\varepsilon$ and $\mu$.
We can then plot the values of $a$, $\rho$ or $\lambda$ in the
$(\varepsilon,\mu)$ parameter space and visualize the effect of varying
the chemical potential (i.e. changing the concentration) or the
adhesion energy (i.e. changing either the adsorbing protein or the
surface
\cite{Ostuni'03,Martins'03})
for a given system.\footnote{
  The chemical potential as defined by Eq. (\ref{mu_d}) also depends
  through $\mu_0$ on $\varepsilon$ as well as $k_\mrm{d}$ and
  $k_\mrm{a}$ which probably depend on $\varepsilon$ as interpreted
  above, but a pure $\ln c$ vs.\ $\varepsilon$ plot can be recovered
  from the $\mu_0 + \ln c$ vs.\ $\varepsilon$ results by a
  transformation of the coordinates.
}
The change of sign of $a$ in the $(\varepsilon,\mu)$-space corresponds,
as shown in Eq. (\ref{a0}) to the curve $\mu=\ln(-\fr{\varepsilon}{M})$
(Fig.\ \ref{afig}). This result is independent of the choice of
$\{l_i\}$-s. An examination of the unit surface coverage $\lambda$
shows a non-monotonic $\mu$ (or $\ln c$) vs.\ $\lambda$ dependence
(Fig.\ \ref{RLfig} b, d, and e) characterized by a maximum and a
minimum, while the total surface density $\rho$ demonstrates similar
unexpected non-monotonic behavior exhibiting a maximum in $\varepsilon$
(Fig.\ \ref{RLfig} a, c, and e). Experimental evidences for this latter
behavior can be found in Ref.\ \cite{Brunner'03}, where the saturation
level of the adsorbed amount of various proteins (HSA, hFb) showed a
non-monotonic dependence on the grafting ratio of PEG chains on TiO$_2$
surfaces.


\subsection{Non-linear potentials}

For potentials with non-linear footprint size dependence, the
equilibrium state density distribution can be calculated numerically
just as easily as for the preceding linear case. The solution, on the
other hand, is no longer a simple exponential distribution and the
potential cannot be as easily parameterized in a physically meaningful
manner.


\section{Dynamics}
\label{secdyn}

In the following we introduce two limiting cases of a general dynamics,
both of which minimize the free energy density (\ref{phi}). In these
models we consider thermodynamically consistent rates for the
transitions of a protein between neighboring states and for the
adsorption and desorption of the protein from and to the smallest
footprint state. These rates are derived from a potential for which
each state is separated by a potential barrier (Fig.\ \ref{potfig}).
Consequently, the rate of a protein expanding from state $i$ (for
$1\leq i<M$) to state $i+1$ (presuming there is sufficient adjacent
empty space) is
\beq k_i^+ = \nu_i \e^{-\fr{\Delta E_{i+1,i}}{2 \kT}},
\label{kip}\eeq
while the rate of shrinking from state $i$ (for $1<i\leq M$) to $i-1$
is
\beq k_i^- = \nu_{i-1} \e^{-\fr{\Delta E_{i-1,i}}{2 \kT}},
\label{kim}\eeq
where $\Delta E_{i+1,i} = \varepsilon_{i+1} - \varepsilon_{i}$ and
$\nu_i$ is determined by the height of the energy barrier.
These rates satisfy detailed balance by construction.
The rate of desorption from state 1 is given as
\beqn k_1^- = k_\mrm{d},\eeqn
while the adsorption rate per unit length (into state 1) is
\beqn k_0^+ = k_\mrm{a} c.\eeqn
To complete a dynamical model we must also have information on the
available surface surrounding each protein at any given time as well as
on the diffusion of proteins along the surface.


\subsection{The infinite diffusion limit}

\begin{figure}
\begin{center}
\includegraphics[scale=0.9]{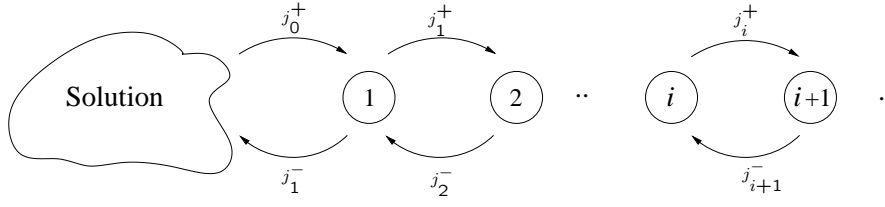}
\end{center}
\caption{
 The complete set of the state density currents $j_i^\pm$.
}
\label{ratefig}
\end{figure}

If the diffusion of adsorbed proteins on the surface is considered
infinitely fast (ideal mixing) we can construct a mean field-like model
where every protein on the surface is presumed to ``feel'' the same
environment. Defining the state density currents (i.e. the number of
transitions per unit time and length)
$j_i^{\pm}$ according to Fig.\ \ref{ratefig}, we can see that every
$j_i^-$ (corresponding to shrinking and desorption) is independent of
the protein environment, whereas every $j_i^+$ (corresponding to
spreading and adsorption) depends on the adjacent free space, which is
considered to be identical for each protein in our mean field-like
model. These currents then have to include the probability of finding a
sufficiently large empty interval to either side. Since we assume the
proteins to be perfectly mixed, the distribution of empty intervals is
exponential with a mean value of $(L-l)/N=(1-\lambda)/\rho$, whence the
probability of finding an interval with a size of at least $x$ is then
\beq P(x) = \e^{-x{\fr{\rho}{1-\lambda}}}
.\label{expp}\eeq
Thus, we can write the state density currents as
\beqn
\left.
\begin{array}{lll}
j_{i+1}^- &= \rho_{i+1} k_{i+1}^- &= \rho_{i+1} k_{i+1}^-\\
j_{i}^+ &= \rho_{i} k_{i}^+ P(l_{i+1}-l_i)
&=\rho_{i} k_{i}^+ \e^{-(l_{i+1}-l_i)\fr{\rho}{1-\lambda}}
\end{array}
\quad \right\} \quad i \ge 1,
\eeqn
and
\beqn
\left.
\begin{array}{lll}
j_{1}^- &= \rho_{1} k_\mrm{d} &= \rho_{1} k_\mrm{d}\\
j_{0}^+ &= \rho_{0} k_\mrm{a} P(l_1)
&=\rho_{0} k_\mrm{a} \e^{-l_1\fr{\rho}{1-\lambda}}
\end{array}
\quad \right\} \quad i=0.
\eeqn
Having derived the currents $j_i^{\pm}$, our system can be completely
described by the following $M$ differential equations:
\beq
\dot{\rho_i} = - ( j_{i}^+ + j^-_{i} ) + (j_{i-1}^+ + j^-_{i+1})
\quad 1\le i \le M
,\label{diffsys}\eeq
which can be readily solved numerically (as an example see Fig.\
\ref{palacsintafig}a).

To verify whether this dynamics converges to the equilibrium state
distribution, characterized by the minimum of the conditional free
energy (\ref{phi}), let us calculate the rate of the change of this
free energy if proteins in state $i$ start to expand to state $i+1$
(for $1\leq i<M$):
\beqn
\fr{\partial \phi}{\partial \rho_{i+1}} -
\fr{\partial \phi}{\partial \rho_{i}}
=
(\varepsilon_{i+1} - \varepsilon_{i}) + \ln
\fr{\rho_{i+1}}{\rho_i}+ \fr{\rho}{1-\lambda} (l_{i+1}-l_i )
.\eeqn
This quantity is identical to
\beqn
\ln \fr{j^-_{i+1}}{j^+_i} = \underbrace{\ln \fr{k^-_{i+1}}{k^+_i}}_{\varepsilon_{i+1} - \varepsilon_{i}} + \ln
\fr{\rho_{i+1}}{\rho_i} + \fr{\rho}{1-\lambda} (l_{i+1}-l_i ),
\eeqn
which means that expansion ($j^+_i$) is faster than shrinking
($j^-_{i+1}$) only if the rate of the free energy change for expansion
is negative. A very similar derivation holds for the relation between
the adsorption ($j^+_0$) and desorption ($j^-_1$) currents. Thus
indeed, the currents between any two states point to a descending
direction on the free energy landscape, driving the system to
equilibrium.

\subsection{The zero diffusion limit}

\begin{figure}
\begin{center}
\includegraphics[scale=0.4]{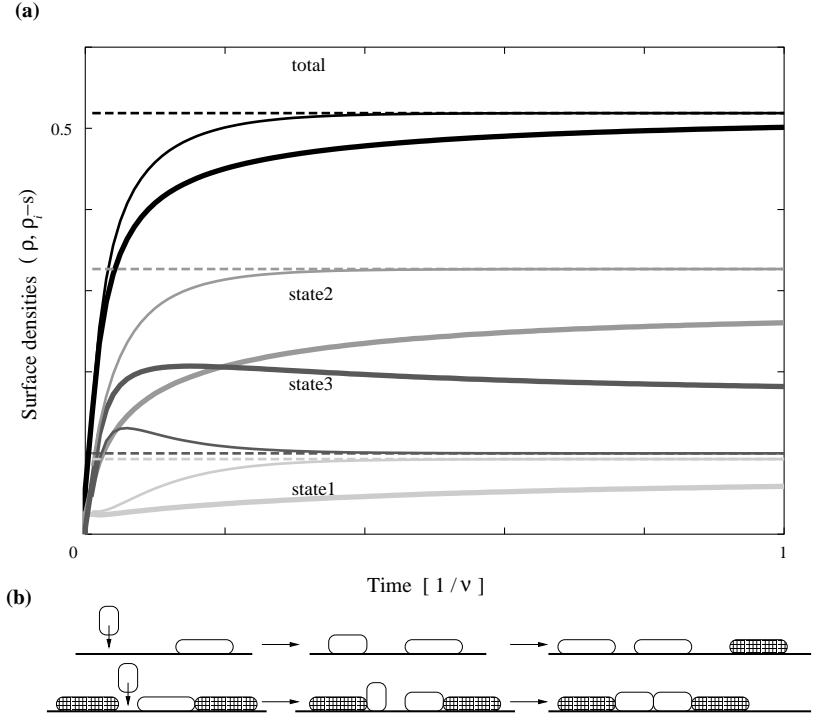}
\end{center}
\caption{
 (a) MC simulation of a protein-surface system with $M=3$ possible
states of sizes $l_1=1$, $l_2=1.5$, and $l_3=2$ with binding energies
$\varepsilon_1=-11.25$ $\kT$,
$\varepsilon_2=-14.0625$ $\kT$, and
$\varepsilon_2=-15$ $\kT$, respectively. The states are separated
by barriers of equal height, characterized by a state independent
$\nu_i=\nu$ prefactor in the corresponding transition rates, as defined
in Eqs.\ (\ref{kip}) and (\ref{kim}). The desorption rate $k_1^-$ is
also given by Eq.\ (\ref{kim}) with the hypothetical $\varepsilon_0$
assumed to be 0. The adsorption rate is set to $k_0^+=\nu c$ with a
dimensionless protein concentration of $c=0.2$. The simulation was
conducted with a system of size $L=200$ and averaged over 10000 runs.
The state densities from the MC simulations (thick lines) as a function
of time (measured in units of $1/\nu$) are plotted together with the
equilibrium distribution (using Eq.\ (\ref{rhoj})) (dashed lines) and
the numerical solution in the infinite diffusion limit (Eq.\
(\ref{diffsys})) (thin lines). The gray levels (starting from light
gray) correspond to $\rho_1$, $\rho_2$, $\rho_3$, and $\rho$.
 (b) Sketches of different regimes of the dynamics: under low surface
coverage all adsorbed proteins spread out to their energetically most
favorable state (top row), while for high coverage -- assuming a convex
potential -- the equally, but not fully spread out state may be
preferential (bottom row).
}
\label{palacsintafig}
\end{figure}

If on the other hand we assume the proteins adsorbed on the surface
immobile\footnote{
  Limited movement occurs none the less, through ``crawling'', i.e.
  successive spreading and receding of proteins resulting in movement
  of the particles' center of mass.
}
after adsorption, we can no longer construct a mean field model such as
above. Transient effects similar to jamming in RSA begin to play a
significant role, the system relaxes very slowly, and the transient
state distributions can be significantly different from the eventual
equilibrium. To study the zero diffusion behavior of our model we
conducted Monte Carlo (MC) simulations of systems of approximately 100
proteins (system size was fixed at $100 \max\{l_i\}$) several hundred
times and averaged the results. Slow relaxation and large differences
in state distribution under similar total surface densities were
observed (see Fig.\ \ref{palacsintafig}) over a wide range of
parameters. The simulations qualitatively reproduced several known
experimental effects, such as the overshot of the surface density at
high concentrations or hysteresis of the surface density as a function
of the concentration even after large equilibration times. We are
performing experiments which will be analyzed by such MC simulations.

\begin{figure}
\begin{center}
\includegraphics[scale=0.4]{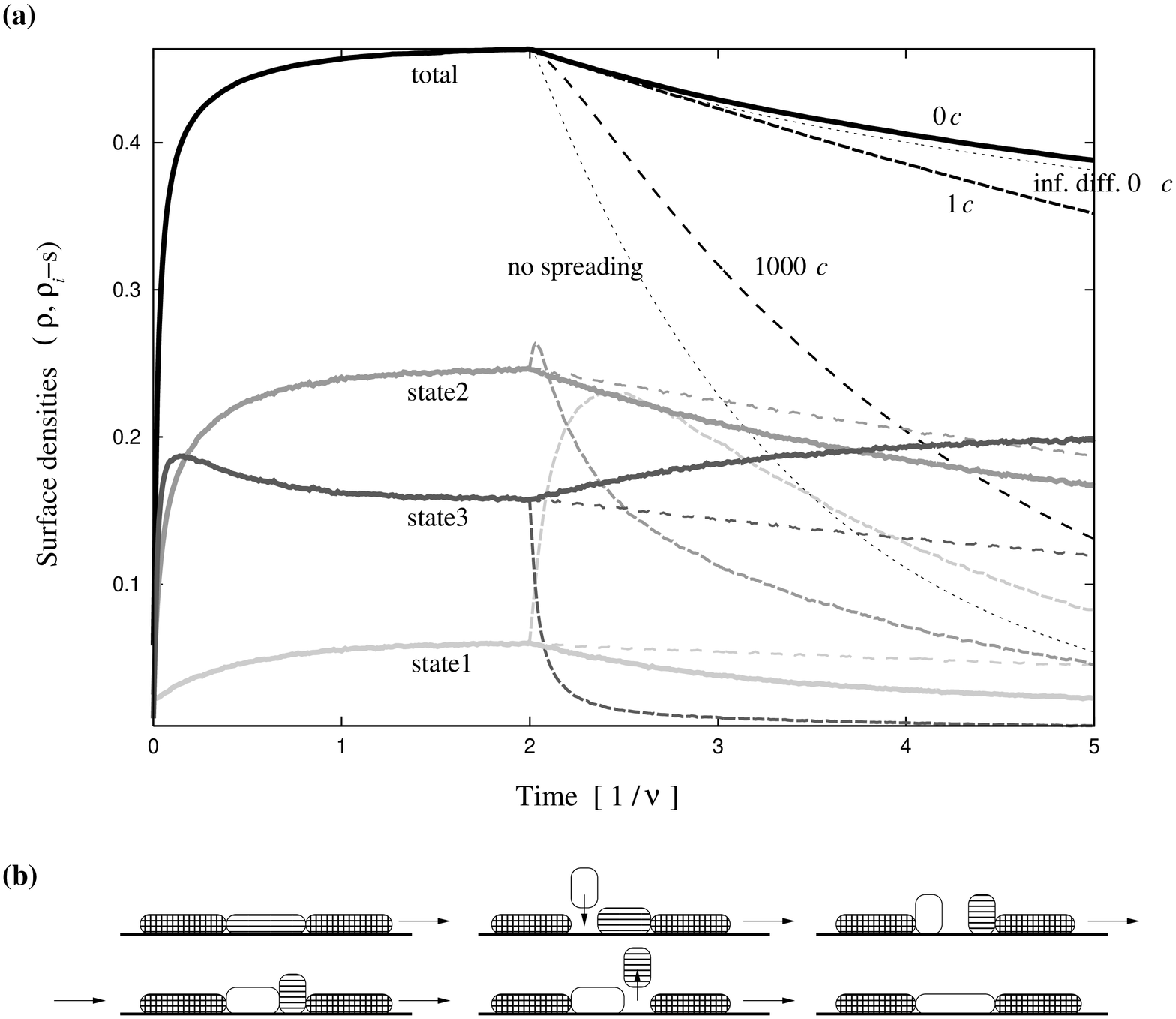}
\end{center}
\caption{
 (a) Plot of the state densities of ``labeled'' proteins from the MC
simulations of the system described in Fig.\ \ref{palacsintafig} (but
averaged over only 2000 runs), using the same color code. As detailed
in the text, the initially solution of labeled proteins (with
concentration $c=0.2$) was replaced at time $2/\nu$ by either a buffer
solution (solid lines), or an ``unlabeled'' solution of the same
concentration (densely dashed lines), or a more concentrated unlabeled
solution (sparsely dashed lines). For comparison, we have also plotted
the total surface density $\rho$ in the infinite diffusion limit with
buffer replacement (upper dotted line), as well as in the limit of the
fastest exchange (i.e., when no spreading of the labeled proteins is
allowed) (lower dotted line).
 (b) Caricature of the ``ratcheting'' mechanism of protein exchange.
}
\label{exfig}
\end{figure}

\subsection{Exchange}

A further consequence of reversibility, aside from the subtle evolution
of the state density, is the possibility of the exchange of the
adsorbed proteins by the proteins of the solution (a phenomenon widely
observed among simple homo-polymers
\cite{Fleer'93}).
To examine exchange we conducted simulations of a hypothetical setup
were ``labeled'' proteins were allowed to adsorb for a specified time
($2/\nu$) and then the solution was replaced by either a buffer
solution (0 $c$ unlabeled proteins), or an ``unlabeled'' solution of
the same concentration (1 $c$ unlabeled proteins), or a more
concentrated unlabeled solution (1000 $c$ unlabeled proteins).
Observing the concentration of labeled proteins on the surface after
the introduction of the new solution (Fig.\ \ref{exfig}), we can see
that the density of labeled proteins decreases faster as the
concentration of unlabeled proteins is increased.

Further examining the various state densities, we see that this
increase can be attributed to the labeled proteins being ``ratcheted''
up to smaller footprint states and eventually expelled from the surface
by the newly adsorbed unlabeled proteins
\cite{Ball'03}.
This is most obvious in case of the replacement with the most
concentrated unlabeled solution (1000 $c$), where a large maximum of
the surface density of state 1 ($\rho_1$) appears near time $2.5/\nu$.
Since desorption occurs exclusively from this state, the inflection
point (i.e., the maximal slope) of the total surface density ($\rho$)
near the same time is another manifestation of the same ratcheting
phenomenon. Because this inflection at high concentrations is a clear
sign of exchange and could easily be measured, we are currently setting
up similar experiments with fluorescently labeled proteins.

\section{Discussion}

We have presented a thermodynamically consistent model of protein
adsorption possessing an equilibrium state. The existence of which
facilitates the understanding and prediction of long-term effects and
also offers novel predictions for the dependence of the equilibrium
surface density ($\rho$) and surface coverage ($\lambda$) on the
binding conditions and the protein concentration.
We have also predicted the existence of a well defined transition
between predominantly compact and expanded states in the adsorbed layer
as a function of either the concentration or the binding conditions.

Introducing dynamics, we derived an analytically tractable mean field
model, which is equivalent to the limit of infinite surface diffusion
of the adsorbed proteins. We have also performed direct MC simulations
of the zero diffusion dynamics, and have been able to qualitatively
reproduce several experimental phenomena. The comparison between the
mean field model and the MC simulations underpins the importance of RSA
like surface frustration in the slowing down of the adsorption
dynamics. The MC simulations have also led us to the idea of a novel
experimental method for the study of protein exchange.


\appendix

\section{Appendix: Configurations on a continuous line}

Starting from a continuous model where each adsorbed protein is located
along a continuous line (1D surface) of adsorption, approximation
(\ref{apC}) can be shown to be exact in the
$\delta\to 0\enskip(n_0\to\infty)$ limit.
To arrive at this result, directly the continuous configuration volume
$\mathcal{C}(\{n_i\})$ has to be derived.

Let us first consider only one adsorbed protein ($N=1$) in state $i$.
This protein separates the adsorption surface into two disjunct
intervals,\footnote{
  This separation of intervals (or shielding property) only holds in
  1D, which is one of the main reasons that 2D calculation are far more
  difficult.
}
each of which can have a length of maximum $L-l_i$. For arbitrary $N$,
each of the $N+1$ intervals has a length
\beqn w_j \in [0,L-l],
\quad \mrm{where} \quad
j\in\{1,2,..,N+1\}
\eeqn
with $l$ being the total occupied surface. The intervals $w_j$ must
also satisfy the spatial constraint
\beq \sum_{j=1}^{N+1} w_j = L - l
.\label{sc}\eeq

\begin{figure}
\begin{center}
\includegraphics[scale=0.4]{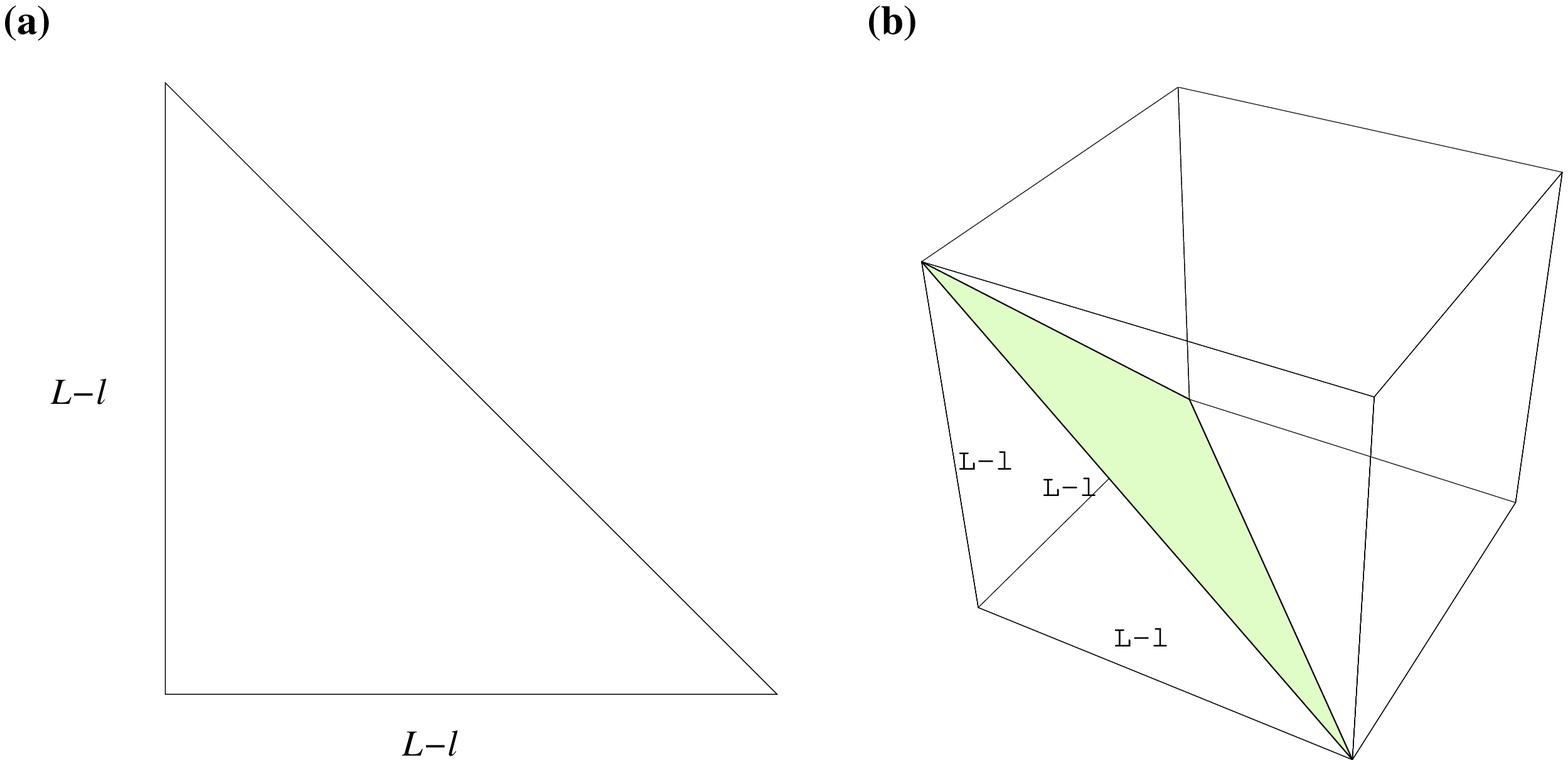}
\end{center}
\caption{
 $\mathcal{C}(\{n_i\})$ for (a) $N=1$ and (b) $N=2$
}
\label{triL}
\end{figure}

This means that the configuration volume $\mathcal{C}(\{n_i\})$ for a
single protein is simply the length of the hypotenuse of a right
triangle with legs of equal length, $L-l$ (Fig.\ \ref{triL}a).

For $N=2$, $\mathcal{C}(\{n_i\})$ is the area of the
equilateral triangle base of the pyramid formed by the origin,
$(L-l,0,0)$, $(0,L-l,0)$, and $(0,0,L-l)$ in the 3D space of $w_1$,
$w_2$ and $w_3$ (Fig.\ \ref{triL}b).
So far we have considered the adsorbed proteins indistinguishable -- by
only considering the lengths and arrangement of the empty intervals,
but not the order of the adsorbed proteins --, consequently this area
has to be multiplied by a factor of two when the proteins are in
different states.

In general for $N>2$ the configuration volume is an $N$-dimensional
hyper-surface in the $N+1$-dimensional space of $w_j$-s defined by the
constraint (\ref{sc}), multiplied by a factor of $\fr{N!}{\prod_{i=1}^M
n_i !}$. This can be written in an integral form as
\beqn
\mathcal{C}(\{n_i\})=\fr{N!}{\prod_{i=1}^M n_i !}\enskip
\int_{-\infty}^{\infty}\prod_{j=1}^{N+1}
\d w_j \enspace \prod_{j=1}^{N+1}
\Theta (w_j)
\enspace\delta(L-l-\sum_{j=1}^{N+1} w_j )
\eeqn
where $\Theta(x)$ is the Heaviside function. Carrying out the
integration with respect to $w_{N+1}$ we have
\beqn
\int_{-\infty}^{\infty}\d w_{N+1}
\enspace\Theta(w_{N+1})
\enspace\delta(L-l-\sum_{j=1}^{N+1} w_j )
=
\Theta (L-l-\sum_{j=1}^{N} w_j)
.\eeqn
Integrating again with respect to
$w_{N}$:
\beqn
\begin{array}{l}
\displaystyle
\int_{-\infty}^{\infty}\d w_{N}
\enspace\Theta(w_{N})
\enspace\Theta(L-l-\sum_{j=1}^{N} w_j)
\\
\displaystyle
=(L-l-\sum_{j=1}^{N-1} w_j)
\enspace\Theta(L-l-\sum_{j=1}^{N-1} w_j).
\end{array}
\eeqn
Now we can see that by integrating with respect to the remaining $N-1$
variables one by one in a similar way, we arrive at
\beqn
\mathcal{C}(\{n_i\})
=
\fr{(L-l)^N}{\prod_{i=1}^M n_i !},
\eeqn
which, divided by the unit volume $\delta^M$, is the same as Eq.\
(\ref{C_d}) with the approximation (\ref{apC}), which can now be
considered exact in the $\delta \to 0$ limit.

\ack

This work was supported by the Hungarian National Science Foundation
(Grant No. OTKA F043756), a Marie Curie European Reintegration Grant
(No. 505969), and the Swiss Federal Institute of Technology.


\end{document}